\documentclass[journal]{IEEEtran}

\IEEEoverridecommandlockouts                              

\usepackage{enumitem}

\usepackage{color}
\usepackage{multirow}
\usepackage{cite}
\usepackage{balance}
\usepackage{amssymb}

\usepackage{soul}
\usepackage{pifont}
\usepackage{url}

\usepackage[pdftex]{graphicx}
\usepackage[table,xcdraw]{xcolor}
\DeclareGraphicsExtensions{.pdf,.jpeg,.png,.JPG,}
\usepackage{graphicx}

\DeclareGraphicsExtensions{.eps}
\DeclareGraphicsExtensions{.tif}
\usepackage{epstopdf}

\graphicspath{{figs/}}
\usepackage{epstopdf}
\usepackage{booktabs}

\usepackage{amsmath}
\usepackage{cleveref}
\ifCLASSOPTIONcompsoc
    \usepackage[caption=false, font=normalsize, labelfont=sf, textfont=sf]{subfig}
\else
\usepackage[caption=false, font=footnotesize]{subfig}
\fi

\usepackage{algorithm,algorithmic}
\begin{document}

\title{Bilateral Market for Distribution-level Coordination of Flexible Resources using Volttron}

\author{Mohammad~Ostadijafari,
        Juan~Carlos~Bedoya,
        Anamika~Dubey,
        and~Chen-Ching~Liu,
\thanks{This work has been supported in part by the U.S. Department of Energy (DOE) under Award DE-0E00840.}
\thanks{M. Ostadijafari and A. Dubey are with the School of Electrical Engineering and Computer Science, Washington State University (WSU), Pullman, WA 99164. e-mail: m.ostadijafari@wsu.edu, anamika.dubey@wsu.edu}
\thanks{J. C. Bedoya and C-C. Liu are with Bradley Department of Electrical and Computer Engineering at Virginia Polytechnic Institute and State University, Blacksburg, VA 24060. e-mail: bedojuan@vt.edu, ccliu@vt.edu.}
}

\IEEEoverridecommandlockouts

\maketitle
\IEEEpubidadjcol
\begin{abstract}
Increasing penetrations of distributed energy resources (DERs) and responsive loads (RLs) in the electric power distribution systems calls for a mechanism for joint supply-demand coordination. Recently, several transactive/bilateral coordination mechanisms have been proposed for the distribution-level coordination of flexible resources. Implementing a transactive market coordination approach requires a secure, reliable, and computationally efficient multi-agent platform. An example of such a platform is  VOLTTRON, developed by the Pacific Northwest National Laboratories (PNNL). The VOLTTRON platform allows the market actors to exchange information and execute proper control actions in a decentralized way. This paper aims to provide a proof-of-concept of the transactive market coordination approach via a small-scale demonstration on the VOLTTRON platform. The steps needed to implement the proposed market architecture using virtual machines and VOLTTRON are thoroughly described, and illustrative examples are provided to show the market-clearing process for different scenarios.

\end{abstract}

\IEEEpeerreviewmaketitle

\section{Introduction}
The increasing penetrations of flexible/responsive loads (RLs) and distributed energy resources (DERs) in the power distribution systems calls for mechanisms to coordinate their trading activities \cite{ostadijafari2020aggregation}. In related literature, numerous centralized and transactive/bilateral mechanisms have been proposed for joint supply-demand coordination \cite{bedoya2019bilateral,5524053,juan2019Decentralized,9247176}. Lately, transactive energy (TE) methods have gained significant attention. TE is defined as a set of economic and control mechanisms to enhance the grid's reliability and efficiency \cite{bedoya2019bilateral,melton2013gridwise}. Different aspects of TE, such as contract design, computational framework development, and market design, are studied in the related literature \cite{8944236,melton2013gridwise,divshali2017multi}.  Likewise, we also proposed a bilateral market for distribution-level coordination of flexible resources (e.g., RL and DERs) \cite{juan2019Decentralized}. The proposed approach allows for bilateral power transactions among market participants that are executed and settled in a fully-decentralized manner. 

The ability to scale for many market participants is one of the main attractions of the transactive market coordination methods, such as the one we previously proposed \cite{juan2019Decentralized}. In transactive methods, the market-clearing algorithms are individually executed by the market participants without the need for a central coordinator or a master node. Thus, it is possible to include many agents as long as they can communicate and bilaterally exchange price-power signals. However, the implementation of such architecture calls for a secure and reliable agent-based platform to allow for the continuous exchange of information among the market's actors. In such a platform, market actors locally perform appropriate control actions based on other agents' provided information \cite{9131248}. 

Several multi-agent platforms (e.g., JADE, ZEUS, Mortar.io, VOLTTRON, etc.) are available in the related literature. Among these platforms, VOLTTRON is specifically designed to be utilized for the smart grid applications, and it is compatible with the current smart grid communication protocols \cite{luo2017investigate}. VOLTTRON is an agent-based, open-source software platform developed by Pacific Northwest National Laboratory (PNNL). Lately, this platform has been used to implement and demonstrate different applications by the power community. For example, authors in \cite{mirakhorli2017open} propose a residential building energy management system in which several simulation agents are designed to model the load consumers in buildings considering grid signal and occupants behavior. In their work, authors employ the VOLTTRON platform to model, control, and monitor major consumers of electricity within the residential buildings and develop use cases for their control to manage grid integration issues. Authors in \cite{haack2013volttron} illustrate the capability of VOLTTRON in coordinating the electric vehicle charging with home energy usage. Reference \cite{chinde2016volttron} utilizes the VOLTTRON platform to control the temperature in a multi-zone commercial building and buildings' water heaters. In \cite{luo2017investigate}, a distributed algorithm is used to solve the economic dispatch (ED) problem, where the VOLTTRON platform is used as a central coordinator to model the interactions among multiple consumers and prosumers. In \cite{xie2018new}, a transactive approach for the energy market is implemented on VOLTTRON.  

This paper aims to provide a proof-of-concept of the bilateral market coordination approach proposed in \cite{juan2019Decentralized} by implementing it using the VOLTTRON platform. To do so, we implement a small-scale test case for the proposed bilateral market coordination approach on the VOLTTRON platform. The implementation uses VOLTTRON as a message bus between the communicating market actors, which are modeled by virtual machines (VMs). This implementation shows how agents in this market bilaterally communicate, exchange price-power signals, and perform proper control actions locally. We also provide several examples to show the market-clearing process in the proposed bilateral coordination approach for different scenarios.

The rest of the paper is organized as follows: Section \ref{sec:2} briefly describes the structure of the bilateral energy market. Section \ref{sec:3} explains the implementation of the market in VOLTTRON. Section \ref{sec:4} provides examples of market-clearing process in different scenarios. Finally, Section \ref{sec:5} presents concluding remarks.

\vspace{-0.2cm}\section{Bilateral Transactive Retail Electricity Market}
\label{sec:2}
\begin{figure}[b]
		\centering
			\vspace{-0.3cm}
		\includegraphics[width=0.9\linewidth]{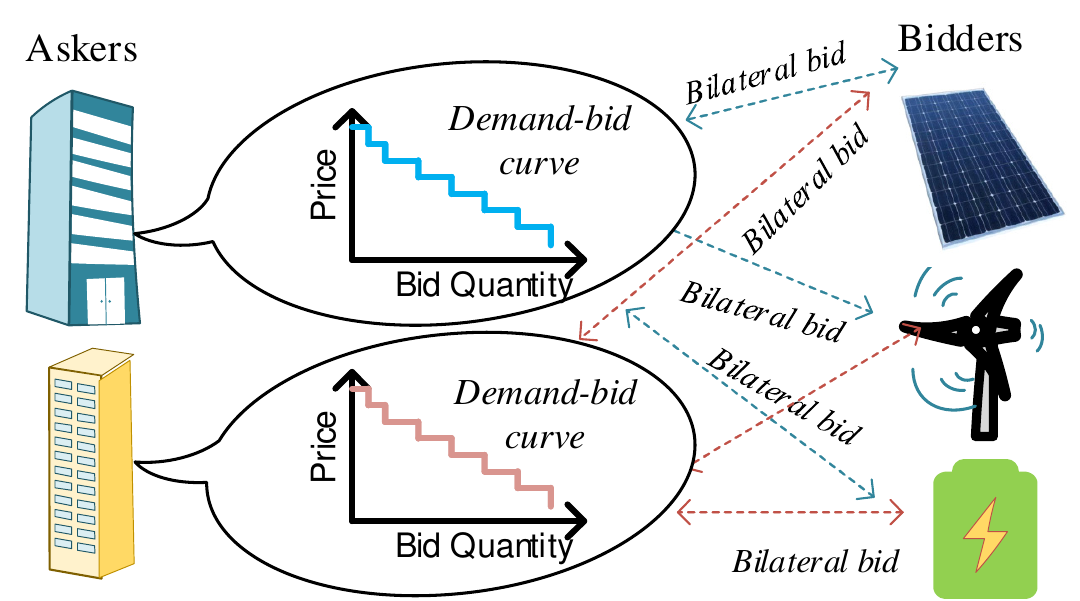}
		\vspace{-0.3cm}
		\caption{Bilateral market}
		\label{fig:1}
	\end{figure}

The bilateral market coordination approach adopted in this work is demonstrated in Fig. \ref{fig:1}. The proposed approach consists of bilateral power transaction agreements in a decentralized manner by market participants \cite{juan2019Decentralized}. The actors of the market are the grid’s proactive participants categorized as “asker agents” (e.g., RLs) and “bidder agents” (e.g., DERs)\cite{ghasemi2020multi}. In this framework, each asker reports its power demand to potential bidders (suppliers) using a demand-bid curve. The asker agents individually develop their demand-bid curves using a model predictive control (MPC) algorithm to schedule their power consumption based on the individual power usage constraints/preferences and their willingness-to-pay for the electricity. Next, each bidder individually and independently obtains its bidding strategies using risk minimization criteria for a minimum rate of return margin (RoRM). Specifically, the bidders minimize the inherent risks associated with the available power selling options based on Markowitz Portfolio Optimization (MPO) and pursue a RoRM that restricts the optimization problem to allow a minimum desired income. Bilateral bids are sent to askers’ agents in a “sealed-bid” and transactions are individually cleared by the askers using the second price sealed-bid auction (SPSBA). To maintain power system security, DSO supervises participants’ intended transactions and proposes new transactions that minimally affect participants’ interests when operational security is threatened.  
In what follows, we describe each of these steps.

\vspace{-0.4cm}\subsection{Asker agents' Demand-Bid Curves Construction}
\label{sec:II}
An asker refers to any market participant that demands electricity and can shift/curtail demand to optimize energy consumption. The prime examples of asker agents are buildings with controllable loads \cite{9087686}. The optimal energy management in buildings requires an efficient algorithm to manage buildings' flexible loads such as heating, ventilation, and air conditioning (HVAC) systems and battery energy storage systems (BESSs) \cite{8956087}. In what follows, we describe the building thermal load model, MPC-based algorithm to co-schedule the HVAC system and the BESS, and mechanism to generate the price-demand curve for the building loads by leveraging the demand-side flexibility. 


\subsubsection{Building Thermal Load Model} We use a simplified RC model for the building thermal dynamics that incorporates the effects of adjustable temperature set-points \cite{wei2016proactive}. The state-space representation of this model is as follows: 
	\vspace{-0.01cm}
	\begin{equation}\label{eqHVAC1}
	\small
	\boldsymbol{x}^{t+1}=\boldsymbol{A}\boldsymbol{x}^{t}+\boldsymbol{B}\boldsymbol{u}^{t}+\boldsymbol{E}\boldsymbol{d}^{t}
, \hspace{0.2 cm} \boldsymbol{y}^t=\boldsymbol{C}\boldsymbol{x}^t
	\end{equation}
	where, $\boldsymbol{d}^{t}$ is the vector of environmental disturbance; $\boldsymbol{A}$, $\boldsymbol{B}$, $\boldsymbol{C}$ and $\boldsymbol{E}$ time-invariant building parameters (see \cite{8956087,wei2016proactive,9184141} for details). For the time $t$, $\boldsymbol{x}^{t}$ is the state vector representing the temperature of the walls and rooms; $\boldsymbol{u}^t$ is the vector of input variables representing air mass flow into each thermal zone; $\boldsymbol{y}^{t}$ is the output vector representing rooms’ temperature. Then, based on the building thermal load model,  the HVAC system power consumption is calculated as the following:
	
		\vspace{-0.1cm}
		\begin{small}
    \begin{equation}\label{eqHVAC2}
	P_{H}^t=c_1(\boldsymbol{u}^t)^3+c_2(\boldsymbol{u}^t)^2+c_3\boldsymbol{u}^t+c_4
	\end{equation}
	\end{small}\noindent where, $P_{H}^t$ is the HVAC cooling system power consumption, and  $c_1, c_2, c_3$ and $c_4$ are constants \cite{wei2016proactive}.
\subsubsection{BESS}
The dynamics for the BESS is formulated in (\ref{eqbat1})-(\ref{eqbat3}) similarly as in \cite{wei2016proactive}. 
\vspace{-0.1cm}
    \begin{equation}\label{eqbat1}
    \small
	{SOC}^{t}=(1-\eta)SOC^{t-1}+\rho \frac{P^{t}_{c,d}}{Q_{bat}}\tau
	\end{equation}
	\vspace{-0.5cm}
	\begin{equation}\label{eqbat2}
	\small
	E^{-}\leqslant {SOC}^{t} \leqslant E^{+}
	\end{equation}
	\vspace{-0.5cm}
	\begin{equation}\label{eqbat3}
	\small
	-d_r\leqslant P^{t}_{c,d} \leqslant c_r 
	\end{equation}

\noindent The energy of BESS is updated based on (\ref{eqbat1}) where ${SOC}^{t}$ is the SOC of BESS, and $P^{t}_{c,d}$ is charging ($P^{t}_{c,d}>0$) or discharging ($P^{t}_{c,d}<0$) power of the BESS at sampling time $t$. The energy decay rate, round-trip efficiency, capacity of the BESS and length of the time-step are shown by $\eta$, $\rho$, $Q_{bat}$ and $\tau$, respectively. Constraints (\ref{eqbat2}) and (\ref{eqbat3}) bound the energy and charging/discharging limits of BESS, respectively, where, $E^{+}$ and $E^{-}$ are the upper and lower limits of energy, respectively;  $d_r$ is the maximum discharge rate and $c_r$ is the maximum charge rate. 

\subsubsection{MPC-Based Building Energy Scheduling Algorithm} 
 The intelligent controller in each building aims at co-schedule the HVAC system with the BESS such that the net cost of transacted energy for a specified time is optimized, and the desired comfort-level is met for building's occupants \cite{8956087}. The problem is formulated as follows: 
	\vspace{-0.1cm}
    \begin{equation}\label{eqMPC1}\small
	\underset{\boldsymbol{u}^t, P_{c,d}^t}{Min}\sum^{t+W-1}_{k=t}{{\lambda}^t.P_{T}^t}
	\end{equation}
\vspace{-0.1cm}	 Subject to:	\vspace{-0.25cm}
    \begin{equation}\label{eqMPC2}\small
    	P_{T}^t=P_{H}^t+P_{c,d}^t, \ P_{T}^t \geq 0
	\end{equation}
    \vspace{-0.3cm}
    \begin{equation}\label{eqMPC3}\small
\underline{\boldsymbol{u}}\leq \boldsymbol{u}^t \leq \overline{\boldsymbol{u}}, \hspace{0.2 cm} \underline{\boldsymbol{y}}^t\leq \boldsymbol{y}^t \leq \overline{\boldsymbol{y}}^t
	\end{equation}
	\vspace{-0.8cm}
	\begin{align*}
	\text{Constraints (\ref{eqHVAC1})-(\ref{eqbat3})}
	\end{align*}
 where, ${\lambda}^t$ and $P_{T}^t$ are the price of energy and total power consumption of the building at the time $t$, respectively. Equation (\ref{eqMPC2}) states that the total power consumption of the building is due to the HVAC system and the BESS charging/discharging. In constraint  (\ref{eqMPC3}), $\underline{\boldsymbol{u}}$ and $\overline{\boldsymbol{u}}$ are lower and upper limits of the air mass flow, respectively; $\underline{\boldsymbol{y}}^t$ and $\overline{\boldsymbol{y}}^t$ are lower and upper limits of room temperature at time $t$, respectively. 


We summarize the MPC-based energy scheduling algorithm for askers in Algorithm $\ref{algo1}$. The algorithm requires the initial state ($\boldsymbol{x}^0$), and the day-ahead information. Solving (\ref{eqMPC1}) results in deriving optimal mass air flow rate trajectory and BESS charging/discharging trajectory for time $t$ to $\small t+W-1$ (line 2). The first entry of these trajectories, $\boldsymbol{u}^t$ and ${P}_{c,d}^t$, are implemented to control the HVAC system and BESS operation (line 3). Then, the current system states are measured and used as the initial values in the next iteration (line-4). Finally, the sampling time and prediction window is advanced by one time-step (line 5), and the algorithm will be applied again.

\setlength{\textfloatsep}{0pt}
\begin{algorithm}[t]
\caption{\small  MPC for the HVAC system}
    \label{algo1}
    \begin{algorithmic}[1]
    \footnotesize
    \renewcommand{\algorithmicrequire}{\textbf{Given:}} 
    \REQUIRE  $\boldsymbol{d}^t, \underline{\boldsymbol{y}}^t, \overline{\boldsymbol{y}}^t$ and $\lambda^t$ \ $\forall$ \ $t \in W$ and $\boldsymbol{x}^0$,$\underline{\boldsymbol{u}}, \overline{\boldsymbol{u}}$
    \FOR {Each $k$}
    \STATE Solve (\ref{eqMPC1})  $\Rightarrow$ $[\boldsymbol{u}^t,\boldsymbol{u}^{t+1}, ...,\boldsymbol{u}^{t+W-1}]$ and $\small[{P}_{c,d}^t, ...,{P}_{c,d}^{t+W-1}]$
    \STATE Apply $\boldsymbol{u}^t$ and ${P}_{c,d}^t$ to control the system  \\
     \STATE  $(\boldsymbol{x}^0|k+1)\leftarrow (\boldsymbol{x}^{t+1}|k)$  
    \STATE  $k\leftarrow k+1$, \hspace{0.2cm} $t:t+W-1\leftarrow t+1:t+W$
 \ENDFOR   
\end{algorithmic}
\end{algorithm}
\subsubsection{Price-Demand Curve Generation}
The price-sensitive demand bid curve is a set of pairs of electricity demand and price forecasts that show the willingness of the individual buildings to purchase a certain volume of electricity based on the corresponding price at each sampling time\cite{wei2016proactive}. 
The approach for price-demand curve generation is summarized in Algorithm $\ref{algo2}$. Specifically, the range of the energy price forecasts $[\underline{\lambda}-\overline{\lambda}]$ is divided into $S$ segments, and a price increment step, ${\lambda}_{inc}$, is defined (line 1). At each iteration, the electricity price for time $t$ is set to $\lambda^t$ while keeping price forecasts for the prices for the rest of the times fixed (lines 3). For each price point, Algorithm $\ref{algo1}$ is executed to obtain the total power demand of the building for the current sampling time (line 4).  
The electricity price $\lambda^t$, and the corresponding power demand ($P_T^t$), constitutes a new data point for the price-demand curve (line 5). The obtained price-demand pairs form the price-demand curve for the current time step.
\setlength{\textfloatsep}{10pt}
\begin{algorithm}[b]
\caption{\small Price-demand curve generation} \label{algo2}
\begin{algorithmic}[1]
\footnotesize
\renewcommand{\algorithmicrequire}{\textbf{Given:}}
\REQUIRE $\underline{\lambda}$ and  $\overline{\lambda}$
\STATE Divide the range $[\underline{\lambda}-\overline{\lambda}]$ to $S$ steps  and define ${\lambda}_{inc}=(\overline{\lambda}-\underline{\lambda})/(S-1)$
\FOR{$i=1:S$}
\STATE {$\lambda^t=\underline{\lambda}+{\lambda}_{inc} (i-1)$}\\
\STATE Do Algorithm $\ref{algo1}$ and obtain $P_{T}^t$
\STATE Store $P_{T}^t$ and $\lambda^t$ as the abscissa and ordinate of a price-demand pair.
\ENDFOR
\end{algorithmic}
\end{algorithm}

 \vspace{-0.25cm}\subsection{Bidding Portfolio Optimization}
 \label{sec:2B} In the bidding portfolio optimization problem, each supplier (bidder agent) selects a set of $L$ askers (neighboring nodes) to carry out the power transactions. The criteria to define the percentage of power capacity ($w_i$ for $i=1,...,L$) to be bilaterally transacted with each asker is determined by solving an MPO problem that minimizes the investment portfolio volatility and satisfies a RoRM. The resulting MPO is formulated as a convex optimization problem as follows:
\begin{equation} \label{eq:D1}
\small
\underset{w_{g,l}}{Min} \ w^T_g .\Sigma_{g}. w_g
\end{equation}
\vspace{-0.2cm}
Subject to: \vspace{-0.02cm}\begin{equation}\label{eq:D2}
\small \sum^{L}_{l=1}\overset{-}{r}_{g,l} w_{g,l} \geq RoRM_g
\end{equation}
\vspace{-0.5cm}\begin{equation}\label{eq:D3}
\small \sum^{L}_{l=1}w_{g,l}=1
\end{equation}
\vspace{-0.3cm}\begin{equation}\label{eq:D4}
\small 0\leq w_{g,l}\leq 1 \  \forall \ l=1,...,L
\end{equation}
The objective function in (\ref{eq:D1}) represents portfolio variance. Random variable ${r}_{g,l}$ represents the rate of return that bidder $g$ would obtain by selling its power to asker $l$. $\overset{-}{r}_{g,l}$ is the expected value of ${r}_{g,l}$. Matrix $\Sigma$ correspond to the covariance matrix of the portfolio, and it is calculated as $L$ investments in (\ref{eq:D5}).
\begin{equation} \label{eq:D5}\small
\Sigma =\begin{bmatrix}
\hspace{-0.9cm} E \left[ \left( r_{g,1}-\bar{r}_{g,1} \right)^2 \right] &\hspace{-2.2cm} ... &\hspace{-1.3cm} E \left[ \left( r_{g,1}-\bar{r}_{g,1} \right) \left( r_{g,L}-\bar{r}_{g,L} \right) \right] \\
\vdots &\hspace{-1cm} \ddots & \vdots \\
E \left[ \left( r_{g,L}-\bar{r}_{g,L} \right) \left( r_{g,1}-\bar{r}_{g,1} \right) \right] & ... &  E \left[ ( r_{g,L}-\bar{r}_{g,L} \right)^2 ]
\end{bmatrix}
\end{equation}

We need to characterize of the behavior of the random variable ${r}_{g,l}$ to calculate the expected values and correlation entries of $\Sigma$ required to solve the bidder's problem in (9)-(12). First, (\ref{eq:D6})-(\ref{eq:D7}) are used to normalize the $L$ price-demand curves with respect to the corresponding generation capacity ($C_g$) and the bilateral bid prices (${P}_{g,l}$). Next, $N$ sample points (${q}_{g,l,n}$ , ${p}_{g,l,n}$) are obtained, where $n=1,...,N$. 
\begin{equation} \label{eq:D6}\small\vspace{-0.1cm}
{q}_{g,l,n}=\frac{Q_{l,n}}{C_g}, \  \forall \ n=1,...,N \ \text{and}\ l=1,...,L
\end{equation}
\begin{equation} \label{eq:D7}\small\vspace{-0.01cm}
{p}_{g,l,n}=\frac{P_{l,n}}{P_{g,l}}, \  \forall \ n=1,...,N \ \text{and}\ l=1,...,L
\end{equation}
where, $ Q_{l,n}$ and $P_{l,n}$ represent abscissa and ordinate of the point $n$ of the price-demand curve for asker agent $l$. 

Next, ${r}_{g,l,n}$ is calculated as in (\ref{eq:D8}), the mean value $\overset{-}{r}_{g,l}$ is calculated in (\ref{eq:D9}), and joint moments or entries $(i,j)$ of covariance matrix $\Sigma [(i,j)=1,…,L]$ are calculated in (\ref{eq:D10}).
\begin{equation} \label{eq:D8}\small\vspace{-0.2cm}
{r}_{g,l,n}={p}_{g,l,n}-1
\end{equation}
\begin{equation} \label{eq:D9}\small\vspace{-0.1cm}
\overset{-}{r}_{g,l}=\frac{1}{N}\sum^{N}_{n=1}{r}_{g,l,n}
\end{equation}
\vspace{-0.2cm}\begin{equation} \label{eq:D10}\small
\Sigma_{[i][j]}=\frac{1}{N}\sum^{N}_{n=1}({r}_{g,i,n}-\overset{-}{r}_{g,i})-({r}_{g,j,n}-\overset{-}{r}_{g,j})
\end{equation}

Finally, upon solving the quadratic optimization problem in (9)-(12), the bilateral power bids for bidder $g$ to asker $l$ are calculated using (\ref{eq:D11}).
\begin{equation} \label{eq:D11}\small\vspace{-0.01cm}
{G}_{g,l}={W}_{g,l}.{C}_{g}
\end{equation}

	
\vspace{-0.5cm}\subsection{Simultaneous auction transactions clearing}\label{sec:2c} After sending the bilateral bids to the asker agents by the bidders, asker agents simultaneously and independently clear the power transactions using a second-price sealed-bid auction (SPSBA). Second-price auction is an application of the Vickrey mechanisms \cite{vickrey1961counterspeculation}. To do so, each asker executes the following procedures:
\begin{itemize}[leftmargin=*]
\item {Asker agent sorts the received bids in ascending order of the price to form a non-decreasing aggregated bid-price curve.}
\item {The intersection point of demand and the aggregated bid-price curve is determined. This point defines the power transaction equilibrium. That is, the amount of power to be transacted corresponds to all the bids located on the left of the equilibrium point in the aggregate bid-price curve.}
\item {The clearing price for each of these transactions is determined based on the second price rule. That is, for the transaction between asker $l$ and bidder $g$, asker $l$ runs the auction without considering the contributions of bidder $g$. The price associated with this second auction's equilibrium point determines the transaction price between $l$ and $g$.}
\end{itemize}  

 	\begin{figure*}[t]
		\centering
		\includegraphics[width=0.99\linewidth]{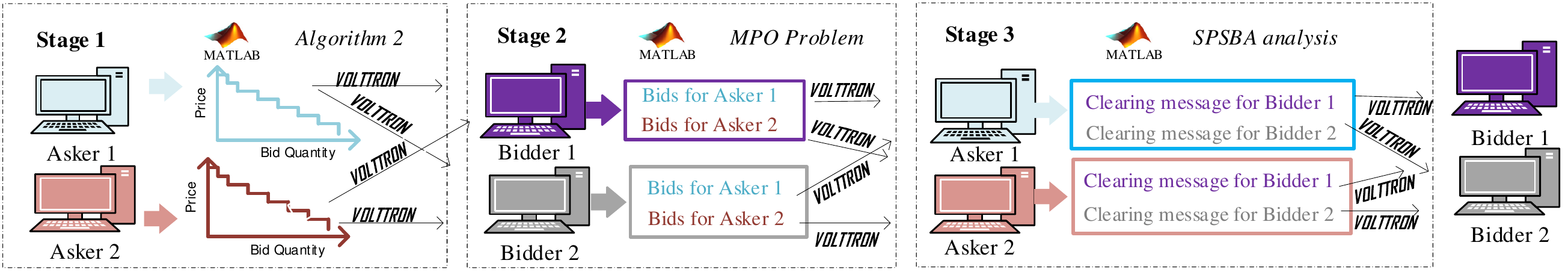}
		\vspace{-0.2cm}
		\caption{The structure of small-scale implementation of the described bilateral market with 4 market actors on VOLTTRON using VMs.}
		\label{fig:3}
		\vspace{-0.4cm}
	\end{figure*}

\vspace{-0.3cm}\section{Implementation on VOLTTRON }\label{sec:3}
Here, we discuss some of the challenges for implementing the proposed market and how the VOLTTRON platform can mitigate these challenges. Then, the implementation of the proposed market using VOLTTRON's agents is described.

\vspace{-0.4cm}\subsection{Implementation Challenges} 
There is a need for a continuous exchange of information among market actors in the proposed bilateral market. Specifically, price-power signals resulting from demand-bid curves,  bidding offers (results of MPO problem), and market-clearing process (results of SPSBA) are exchanged among market actors. For the proper functioning of the proposed bilateral market coordination approach, secure communication must be established among the market actors. First, for an unbiased market, the privacy of the information should be maintained. For example, sending bids from a bidder to multiple askers should not be revealed to other bidders; this prevents other bidders from manipulating their bidding offers to win auctions. Second, communications are prone to cyber-attacks. In this case, the attackers can disrupt the normal market operation by manipulating the exchanged information. For example, attackers can falsify price-demand curves generated by different askers, resulting in a supply-demand imbalance, causing instability in grid operations and cascading failures in extreme cases \cite{soltan}. Similarly, a manipulated high bidding offers can result in denial of participation by askers in the market; hence, poor demand response. These challenges highlight the necessity for a secure and reliable communication platform between different market entities in the proposed architecture. 
 
The VOLTTRON platform helps establish secure communication among the market actors in the proposed framework. The VOLTTRON platform provides a central message bus, allowing the agents (modeled as VOLTTRON agents) to exchange information by publishing and subscribing to topics. VOLTTRON uses VIP protocol to increase the security of the message bus and between VOLTTRON platforms. VIP authorization protects subscribers on the platform from receiving messages from unauthorized agents \cite{lutes2014volttron}. Specifically, the platform owner can limit who can publish to a given topic. This protects subscribers from receiving messages from unauthorized agents.
 
Another critical requirement for implementing the proposed market structure is a platform that is economical and computationally efficient. The transition to the proposed market economically is expensive as market actors need to be equipped with intelligent processors. These processors should also respond to the control signal in the specified time; thus, they need to be computationally efficient. Furthermore, the installation of the communications platform should not add high cost or computational requirements. Fortunately, VOLTTRON does not consume considerable processing resources, and it can run on a single board small computer, such as Raspberry PI \cite{lutes2014volttron}.    

\setlength{\textfloatsep}{10pt}
  \begin{small}
 \begin{algorithm}[b]
 \caption{\small Market implementation on VOLTTRON}
     \label{algo3}
     \begin{algorithmic}[1]
     \footnotesize
     \renewcommand{\algorithmicrequire}{\textbf{Stage 1:}} 
     \REQUIRE Sending demand-bid curves
     \FOR {each askser} 
     \STATE Perform Algorithm \ref{algo2} $\Longrightarrow$ "demand-bid" message\\
     \STATE Send "demand-bid" message to all bidders \\
  \ENDFOR
  \renewcommand{\algorithmicrequire}{\textbf{Stage 2:}} 
     \REQUIRE MPO problem
     \FOR {each bidder} 
     \STATE Solve MPO problem $\Longrightarrow$ "bid-offer" message\\
     \STATE Send each of "bid-offer" messages to the corresponding asker \\
  \ENDFOR
   \renewcommand{\algorithmicrequire}{\textbf{Stage 3:}} 
     \REQUIRE SPSBA analysis
     \FOR {each asker} 
     \STATE Perform SPSBA analysis $\Longrightarrow$ "market-clearing" message\\
     \STATE Send "market-clearing" messages to the corresponding bidder \\
  \ENDFOR
 \end{algorithmic} 
 \end{algorithm} %
 \end{small}
 
 \vspace{-0.5cm}\subsection{Implementation of the Transactive Demand-Supply Coordination using VOLTTRON}\label{sec:3-2} 
 The implementation of the proposed transactive demand-supply coordination approach in VOLTTRON is summarized in Algorithm $\ref{algo3}$. Communications are based on modeling the market actors as VOLTTRON nodes, i.e., publisher (sender) nodes and subscriber (receiver) nodes based on the data flow. At each stage of the algorithm, bidder and asker information are collected and forwarded among appropriate VOLTTRON nodes using VOLLTRON agents through a message bus. Specifically, on the publisher node, the publisher agent publishes device data to the message bus. From the message bus, the forwarder agent is used to transfer data to the subscriber nodes. On the subscriber node, the listener agent is used to receive the messages with the specified topics. In this communication approach, the communicating nodes do not communicate directly nor necessarily know each other, but instead, they exchange messages by publishing/subscribing to the message bus based on mutual topics. This topic-based authentication controls the messages that different agents can access. Note that this approach is only one example of implementing transactive mechanisms in VOLTTRON, and there are other methods and agents for such purposes in this platform (see \cite{lutes2014volttron} for details). Next, we describe the different stages of the algorithm.

 In the first stage,  askers construct their demand-bid curves based on Algorithm \ref{algo2}. We name the constructed price-quantity pairs "demand-bid" messages. Askers should send their demand-bid message to other bidders in the market. To do so, each asker uses its VOLTTRON publishing agent to publish demand-bid messages to the message bus. Finally, VOLTTRON's forwarder agent of each asker sends the published demand-bid messages to the bidders.

In the second stage, each bidder receives demand-bid messages by using its VOLTTRON listener agent. Then, bidders perform MPO analysis (see section \ref{sec:2B}), which results in finding the volume of powers and the corresponding bilateral prices to be transacted with each asker. We name these price-quantity pairs "bid-offer" messages. Each bidder should send each of its bidding messages to the corresponding askers. Similar to the first stage, publishers and forwarders agents are used to transfer the bidder's information to the askers using the message bus.

In the third stage, each asker receives all different bidding messages by using its VOLTTRON listener agent. Then, each asker performs SPSBA analysis (see section  \ref{sec:2c}), which results in pairs of values showing the amount of power with the corresponding price that each asker will buy from each bidder; we name these sets of pairs as "market-clearing" messages. Each asker should inform bidders of their corresponding market-clearing messages. Similar to the previous stages, publishers and forwarders agents in VOLTTRON are used to publishing the messages to the message bus and send it to their corresponding bidders. At the end of this stage, the demand-bids and the bid-offers are cleared.


\vspace{-0.3cm}\section{Case Study}\label{sec:4}
\begin{figure}[b]
		\centering
			\vspace{-0.2cm}
		\includegraphics[width=\linewidth]{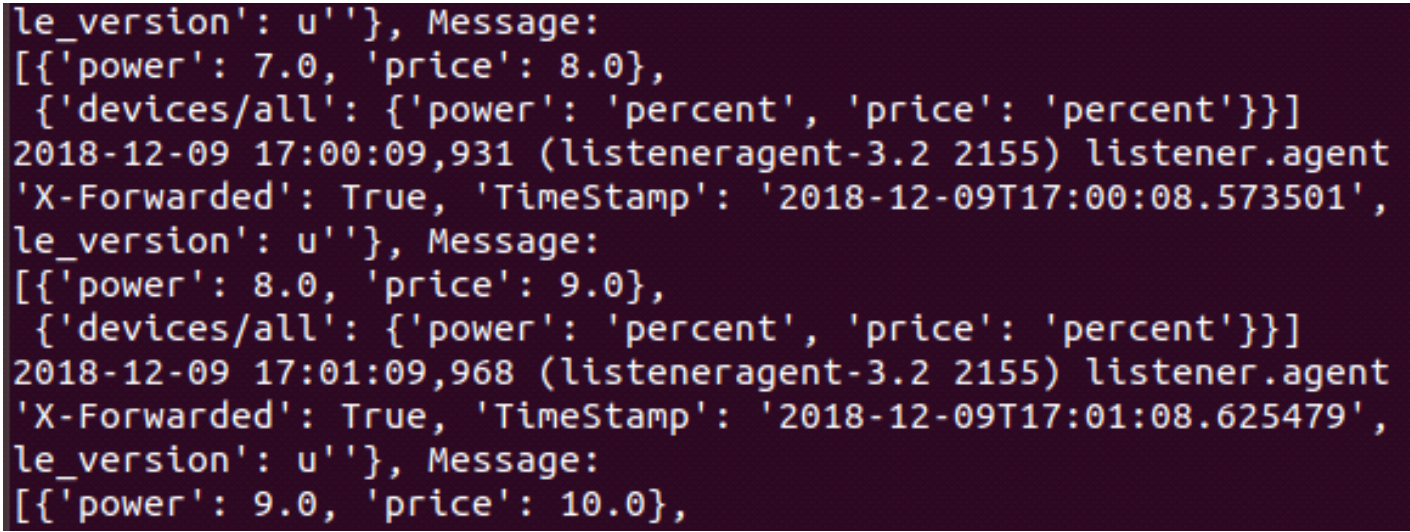}
		\vspace{-0.7cm}
		\caption{Exchange demand-bid messages in VOLTTRON}
		\label{fig:VOLTTRON}
	\end{figure}

Fig. \ref{fig:3} shows the structure of the small-scale implementation of the described bilateral market using 4 market actors (2 askers and 2 bidders) on the VOLTTRON platform. Each market actor is simulated using a Virtual Machine (VM). All VMs works with Linux operating system, and they are installed on a desktop PC with a dual-core i7 3.41 GHz processor, 16 GB of RAM. VOLTTRON is installed in each VM. There is a MATLAB code in each VM, acting as its intelligent agent with the following steps: (1) activating VOLTTRON, (2) extracting the desirable message from VOLTTRON's log file, (3) performing appropriate market analysis based on the extracted information and stage of the Algorithm \ref{algo3} to generate the price-power signals (i.e., messages), (4) saving the message in a compatible format with the VOLLTRON, and (5) publishing the signal in the message bus between the communicating nodes. Fig. \ref{fig:VOLTTRON} shows the zoomed-in view of the VOLTTRON environment for bidder 1 upon receiving a "demand-bid" message from asker 1.

\begin{figure}[b]
     \centering
    \vspace{-0.4cm}
    \subfloat[Auction 1\label{fig:1-1}]{
		\includegraphics[width=0.485\linewidth]{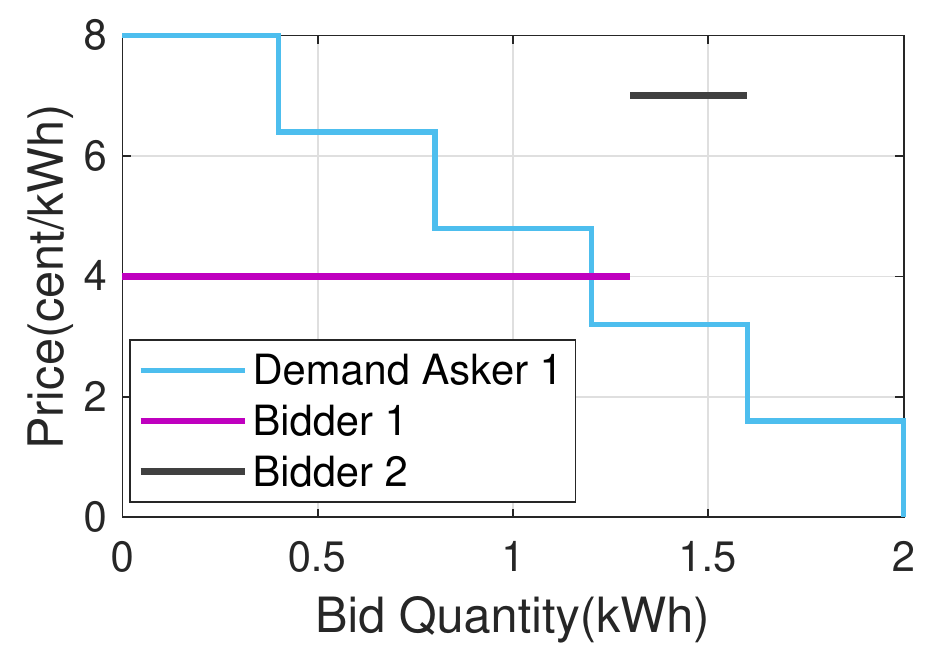}}  
	    \subfloat[Auction 2 \label{fig:1-2}]{
		\includegraphics[width=0.485\linewidth]{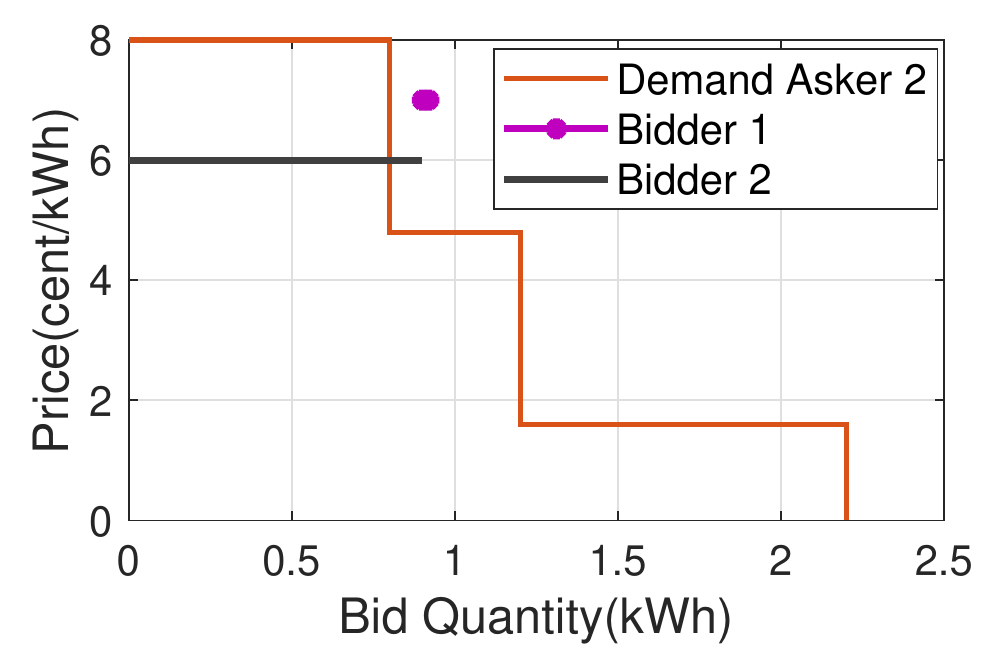}}
  \caption{Scenario 1}
\label{fig:scenario1}
\end{figure}

Next, we illustrate the aforementioned implementation of the transactive demand-supply coordination mechanism on the VOLTTRON platform using three different test scenarios. The price-demand curves of asker 1 and asker 2 are the same in all of these scenarios (see Fig. \ref{fig:scenario1}). Asker 1 requires 2 kW with 1.6 cents/kW of willingness-to-pay. The first 0.4 kW of requested power, for which asker 1 can pay up to 8 cents/kW, can be understood as the minimum power demand required by asker 1. Here, the required electricity power follows ``Normal good" price rules, where higher prices imply less demand \cite{NormGood}. Once the price-demand curves' information is sent to potential bidders, each of them solves the MPO problem (see section \ref{sec:2B}). For each of the three simulation scenarios, it is assumed that bidder 1 has a competitive advantage over bidder 2, which allows bidder 1 to hold lower bidding prices than bidder 2.

For scenario 1, it is assumed that bidder 1 has a lower rate of return than bidder 2. Thus, bidder 1 bids all its available (1.3 kW) capacity to the asker 1 @4.0cent/kW. Bidder 1 does not submit bids to asker 2 as it expects to be fully dispatched in the auction of asker 1. On the other hand, bidder 2, with 1.2 kW of available power, decides to bid 0.9 kW (@6.0 cent/kW) to asker 2 and 0.3 kW (@7.0 cents/kW) to asker 1. Finally, both bidders send their price-power bids to each asker. Askers receive the price-power bids, and they clear the transactions via the SPSBA auction mechanism as detailed in section  \ref{sec:2c}. The market-clearing results are also shown in Fig. \ref{fig:scenario1}. In this scenario, each bidder's risk portfolio strategy allows them to be dispatched at each of the asker's auctions, and the cleared transactions allow supplying both askers power demands.

\begin{figure}[!t]
     \centering
    \subfloat[Auction 1\label{fig:2-1}]{
		\includegraphics[width=0.485\linewidth]{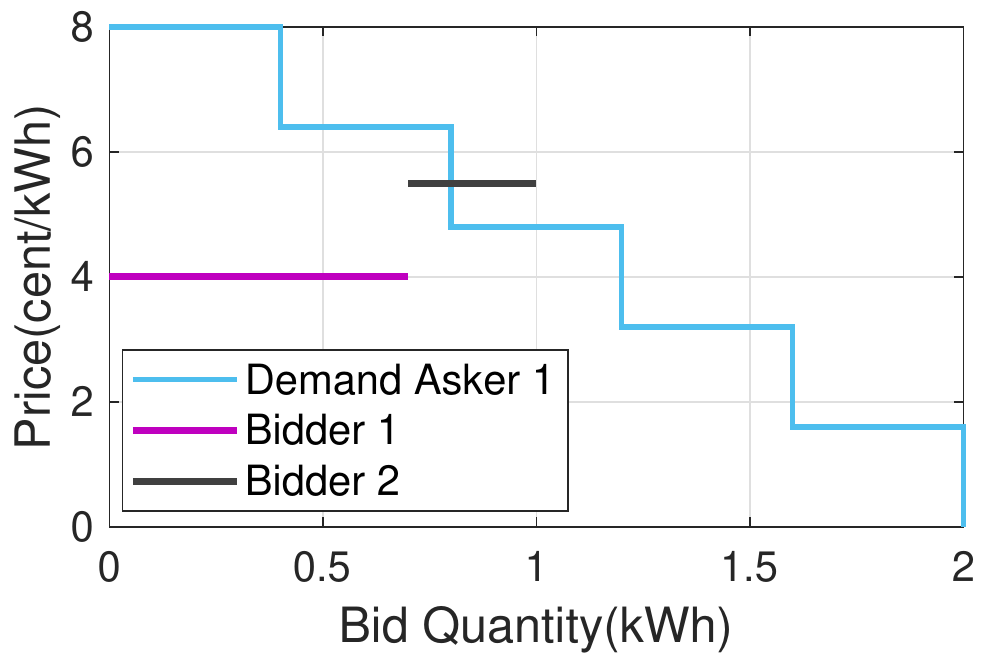}}  
	    \subfloat[Auction 2\label{fig:2-2}]{
		\includegraphics[width=0.485\linewidth]{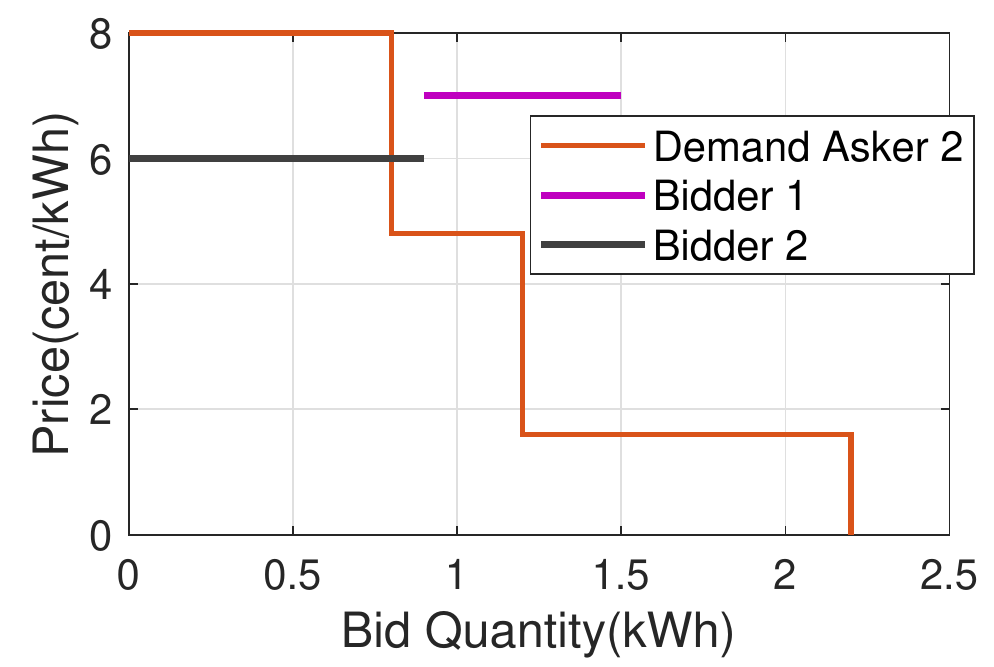}}
  \caption{Scenario 2}
\label{fig:scenario2}
\vspace{-0.2cm}
\end{figure}

In the second scenario, the rate of return for bidder 1 is increased while bidder's 2 rate of return is the same as scenario 1. Thus, bidder 1 now decides to distribute its available power to the auctions carried by both askers. Bidder 1 offers 0.7 kW (@ 4.0 cent/kW) to asker 1 and 0.6 kW (@7.0 cent/kW) to asker 2. On the other hand, bidder 2 offers 0.3 kW (@ 5.5 cent/kW) to asker 1 and 0.9 kW (@6.0 cent/kW) to asker 2. In this case, bidder 1 and bidder 2 are partially dispatched in either asker's auctions. Compared to scenario 1, bidder 1 deploys a larger part of its power portfolio in the auction of asker 2, where it can bid for a higher electricity price. In other words, the bidder relies on the auction, where it is expected to get a better rate of return. As a combined effect of these strategies, bidder 2 has more participation in this scenario; thus, it can obtain a part of the power supply in the auction carried by the asker 1 and its entire power supply for the auction carried by asker 2 (see Fig. \ref{fig:scenario2}).


\begin{figure}[hbt]
     \centering
     \vspace{-0.5cm}
    \subfloat[Auction 1\label{fig:3-1}]{
		\includegraphics[width=0.485\linewidth]{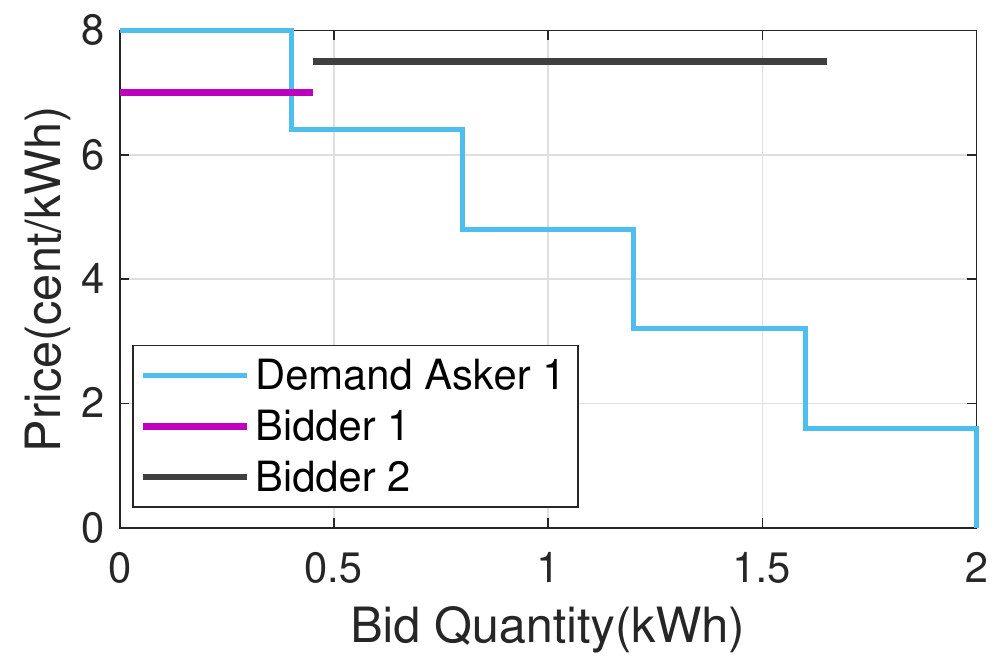}}  
	    \subfloat[Auction 2 \label{fig:3-2}]{
		\includegraphics[width=0.485\linewidth]{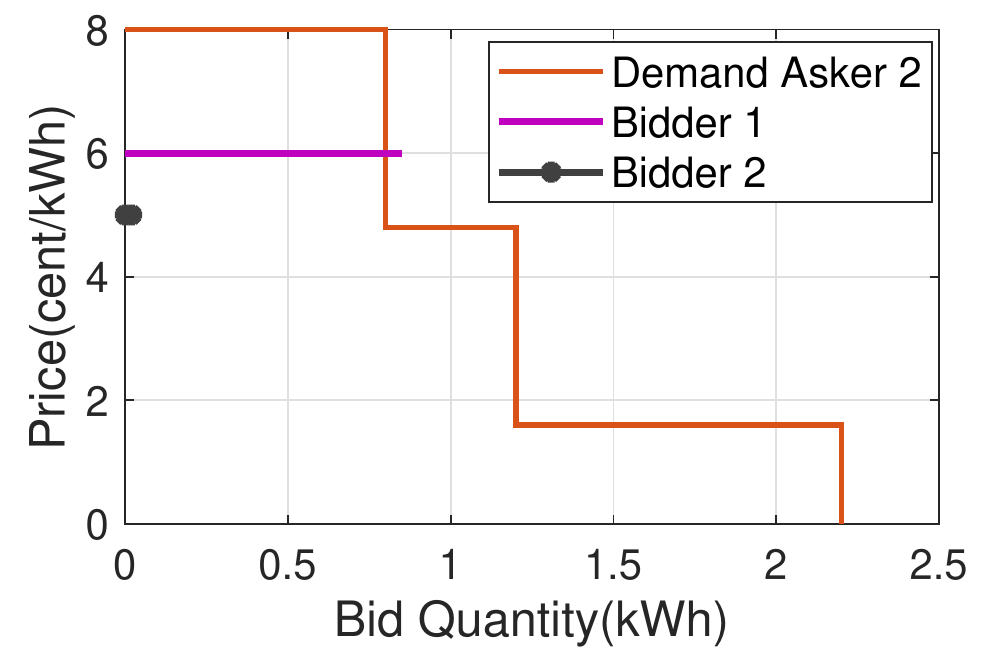}}
  \caption{Scenario 3}
\label{fig:scenario3}
\vspace{-0.3cm}
\end{figure}

In the third scenario, risk assessment conditions for both bidders are increased. Hence, both bidders decide to bid most of their available power to the asker with the most elastic price-demand curve. When a bidder has a higher rate of return expectations, the risk minimization problem leads the bidding portfolio to those assets that include more "flexible" products, i.e., more price-demand elasticity \cite{juan2019Decentralized}. Thus, it is more likely for bidder 1 and bidder 2 to be dispatched in the auction of asker 1.
Bidder 1 and bidder 2 offer 0.45 kW (@7.0 cent/kW) and 1.2 kW (@ 7.5 cent/kW) to asker 1, respectively. For asker 2, only bidder 1 offers at 0.85 kW (@6.0 cent/kW). The market-clearing results are shown in Fig. \ref{fig:scenario3}. The bidder with the highest competitive advantage is dispatched in both auctions. Although risk assessment for bidder 2 is increased for this scenario, the expected rate of return biases is bid to offer all of the available power on asker’s 1 auction. However, given that bidder 1 has a competitive offer, bidder 2 is not dispatched (its competitor has a lower bid price). 

\section{Conclusion}\label{sec:5}
This paper aims at providing a proof-of-concept of a transactive supply-demand coordination approach for the retail electricity market by implementing it using VOLTTRON -- a multi-agent platform for the smart grid application. A small-scale example of the proposed bilateral supply-demand coordination approach is implemented, where market actors simulated using virtual machines (VMs) and exchange information with each other using the VOLTTRON platform. A brief overview of the VOLTTRON and the implementation steps of the bilateral market on this platform are detailed. Finally, with the help of illustrative examples, we showed the practicality of the bilateral market in jointly coordinating the supply-demand in the distribution-level.

\ifCLASSOPTIONcaptionsoff
  \newpage
\fi

	\bibliographystyle{IEEEtran}
	\bibliography{Refrence}

\end{document}